\documentclass[12pt,a4paper]{article}
\usepackage{amsmath}
\usepackage{amssymb}
\usepackage{bm}
\usepackage{geometry}
\usepackage{setspace}
\usepackage{graphicx}
\usepackage{lscape}
\usepackage{verbatim}
\usepackage{natbib}
\usepackage{xcolor}
\usepackage{tabularx}
\usepackage{ragged2e}
\definecolor{winered}{rgb}{0.5,0,0}
\usepackage[bookmarks=true, bookmarksnumbered=true, allbordercolors={1 1 1}]{hyperref}
\hypersetup{
  colorlinks   = true, 
  urlcolor     = blue, 
  linkcolor    = blue, 
  citecolor    = winered,
}
\usepackage[open=true, numbered=true]{bookmark}
\usepackage{float}
\usepackage{fancyhdr}
\usepackage{appendix}
\usepackage{scalefnt}
\usepackage{afterpage}
\usepackage[left]{lineno}
\usepackage{caption}
\usepackage{subcaption}
\usepackage{varioref}
\usepackage{threeparttable}
\usepackage{booktabs}
\usepackage{csquotes}
\usepackage{cleveref}
\usepackage{rotating}
\usepackage{amsfonts}

\makeatletter
\def\blfootnote{\xdef\@thefnmark{}\@footnotetext}
\makeatother

\emergencystretch=\maxdimen
\begin{document}
\title{\LARGE{Taking the Highway or the Green Road? \\Conditional Temperature Forecasts Under Alternative SSP Scenarios}}
\author{Anthoulla Phella{\small \thanks{Corresponding author. \textit{E-mail address}:
anthoulla.phella@glasgow.ac.uk (A. Phella)}} \\ \emph{University of Glasgow}\\
\and Vasco J. Gabriel  \\ \emph{University of Victoria and NIPE} \\
\and Luis F. Martins \\ \emph{ISCTE -- Instituto Universit\'{a}rio de Lisboa and CIMS}}

\date{August 2025}

\maketitle
\begin{abstract}
\noindent 
In this paper, using the Bayesian VAR framework suggested by \cite{Chanetal2025}, we produce conditional temperature forecasts up until 2050, by exploiting both equality and inequality constraints on climate drivers like carbon dioxide or methane emissions. Engaging in a counterfactual scenario analysis by imposing a Shared Socioeconomic Pathways (SSPs) scenario of \enquote{business-as-usual}, with no mitigation and high emissions, we observe that conditional and unconditional forecasts would follow a similar path. Instead, if a high mitigation with low emissions scenario were to be followed, the conditional temperature paths would remain below the unconditional trajectory after 2040, i.e. temperatures increases can potentially slow down in a meaningful way, but the lags for changes in emissions to have an effect are quite substantial.  The latter should be taken into account greatly when designing response policies to climate change. 

\bigskip

\bigskip \medskip

\end{abstract}
\thispagestyle{empty} 

\newpage
\onehalfspace

\section{Introduction}

Forecasting climate variables and, in particular, temperatures is crucial for understanding future environmental conditions and making informed policy decisions. Traditional forecasting approaches primarily rely on statistical or structural models that optimize for the most probable outcomes under given conditions, and often involve an extensive list of assumptions with respect to the relationships across model variables (\citealp{Hasselmann1993}; \citealp{Stockeretal2013}). Meanwhile, in many cases, conducting counterfactual analysis can provide a framework for exploring alternative climate outcomes by conditioning on constraints that may reflect policy targets, physical limits, or hypothetical scenarios. These constraints often take the form of equality conditions—where specific climate drivers such as $CO_2$ emissions or energy use are fixed at predetermined levels by imposing a specific equality. However, given the uncertainty that surrounds the ability and willingness of the world to adapt to climate change and reduce emissions, it can be of interest to allow these constraints to vary within certain thresholds.  Thus, we propose the use of a Bayesian Vector autoregression (VAR) framework, in particular as that is outlined in \citet{Chanetal2025} to produce conditional temperature forecasts by imposing both equality and inequality constraints on climate drivers, focusing on potential paths for a range of greenhouse gases.

VAR models offer a robust approach to multivariate time series analysis, accommodating the endogenous relationships among climate and socioeconomic variables. Unlike univariate models, VAR models do not impose restrictive assumptions on the direction of causality, thus allowing for a comprehensive analysis of the interplay between variables. The efficacy of VAR models in capturing the temporal dynamics of complex systems makes them particularly suitable for climate forecasting \citep{stock1998} and a great complement to the widely used structural energy models, such as Integrated Assessment Models (IAMs) and Earth System Models (ESMs), that can often rely on numerous assumptions about economic growth, energy use, and technological advancements (\citealp{Nordhaus1994}; \citealp{VanVuurenetal2011}). Conditional forecasting involves generating forecasts based on specific assumptions or conditions about the future paths of certain variables. In the context of VAR models, this means projecting the future values of endogenous variables given certain constraints on exogenous variables. Conditional forecasts are valuable for policymakers and researchers as they allow for scenario analysis and the assessment of potential outcomes under different conditions \citep{waggoner1999}. 

Vector Autoregression models have recently been utilized in climate science to forecast temperature changes and other climatic variables. In a study by \citet{nuruzzaman2023}, a VAR model was employed to forecast temperature, rainfall, and cloud coverage for the Jessore region of Bangladesh. While the stationarity of variables was determined using ADF, PP, and KPSS unit root tests, the Granger causality test was used to verify the endogeneity among the variables. The study revealed a trend toward increasing temperature and a trend toward decreasing rainfall and cloud coverage. Similarly, \citet{si2023} developed a large VAR model to forecast three important weather variables for 61 cities across the United States. The study modeled temperature, precipitation, and wind speed as response variables. The VAR model demonstrated its efficacy in capturing the temporal dynamics of these weather variables, providing valuable insights for electricity supply and demand forecasting. 

The application of VAR models in temperature forecasting has proven to be a robust approach for analyzing the dynamic interactions among climatic variables, albeit not without drawbacks. The accuracy of a VAR model heavily depends on the correct specification and selection of appropriate lag lengths. In climate forecasting in particular, determining the optimal lag length can be challenging due to the complex and nonlinear nature of climatic processes. Incorrect lag selection can result in biased estimates and poor forecasting performance \citep{stock1998}. Furthermore, climatic variables often exhibit non-stationary behavior due to long-term trends and seasonal patterns, while VAR models assume that all variables in the system are stationary or can be transformed to achieve stationarity. Failure to properly address non-stationarity can lead to spurious results and unreliable forecasts \citep{johansen1995}, although this is not an issue with the Bayesian framework employed here. Despite these limitations, VAR models remain valuable tools for analyzing the dynamic interactions among multiple climatic variables, as long as researchers are aware of these constraints and apply appropriate techniques to mitigate their impact, ensuring more accurate and reliable climate forecasts.

Taking the aforementioned into consideration, the primary objective of this study is to demonstrate the utility of VAR models in (ex-ante) forecasting temperature changes across different Shared Socioeconomic Pathways (SSPs) scenarios. Traditional forecasting methods rely on ``internal'' model-based assumptions about the evolution of the drivers of the variable of interest. In our setup, we employ ``externally'' validated scenarios to inform our predictions. The SSP framework, developed by the scientific community as part of the Intergovernmental Panel on Climate Change (IPCC) assessments, delineates five distinct pathways that describe potential global developments and their associated emission trajectories. These pathways, ranging from sustainable development (SSP1) to significant challenges to mitigation and adaptation (SSP5), serve as a basis for examining the implications of varying socioeconomic conditions on future climate projections \citep{riahi2017}. By incorporating the emissions outlined by the different SSP scenarios into a multivariate model examining the evolution of key climate variables, this research aims to contribute to the broader understanding of how different socioeconomic pathways influence climatic outcomes and to support the development of effective mitigation and adaptation strategies. Accurate forecasting of temperature variations under different socioeconomic scenarios is paramount for informed policy-making and strategic planning and this study underscores the significance of incorporating advanced econometric methods in climate science to enhance the accuracy and reliability of long-term climate projections.\footnote{We also direct the reader to \citet{HENDRY2023754} for a discussion on whether scenario comparisons can be informative and how inferences about scenario differences depend on the relationships between the conditioning variables.} 

Nevertheless, given the uncertainty surrounding the ability to implement and achieve such scenarios, imposing only strict equality conditions may not always be realistic. Equality constraints specify that certain variables must take on specific values or follow a predetermined path over the forecast horizon. These constraints are often used in policy analysis to simulate the effects of specific interventions or to ensure consistency with known future events. On the other hand, inequality constraints specify that certain variables must lie within a specified range or follow a path that satisfies certain conditions. These constraints are useful for incorporating realistic bounds on variables, such as non-negativity constraints on prices or emissions limits in climate models. 

Inequality constraints allow for more flexible and realistic scenario analysis compared to equality constraints. Therefore, we propose the use of the Bayesian VAR as proposed in \cite{Chanetal2025}, which allows for both multiple equality and inequality constraints. Their closed-from solution makes their method suitable for both conditional forecasts and scenario analysis, in contrast with previous work which has previously engaged in inequality constrained conditional forecasts (see \textit{inter alia}, \citealp{waggoner1999conditional, andersson2010density}). Furthermore, the authors additionally derive the conditional forecasts' distribution in a way which allows the model to handle a large dimensional VAR or a large number of conditioning variables and long forecasts horizons more efficiently. This can be highly relevant for climate applications.\footnote{In practice, this is achieved by presuming the conditional forecasts as time series with missing data and make use of the efficient sampling algorithm proposed in \citet{chan2023high}.}

Indeed, our conditional forecasting framework provides an important extension to the conventional methodologies delivering probabilistic projections of global near-surface temperature. However, these forecasts are typically unconditional, in the sense that they aggregate across model ensembles without explicitly conditioning on alternative external drivers or boundary conditions. A conditional framework addresses this limitation by embedding forecasts within specific assumptions about drivers of temperatures (such as greenhouse gas trajectories), thereby allowing direct exploration of scenario-dependent temperature pathways. This refinement adds substantial value to unconditional approaches: while climate models produce robust ensemble-based probabilities of exceeding thresholds such as 1.5\,°C above pre-industrial levels, conditional forecasting highlights how those probabilities shift under distinct policy or geophysical contingencies. In comparison to existing ensemble-mean approaches, conditional forecasts reduce uncertainty in a transparent manner and improve attribution of near-term anomalies by disentangling forced responses from natural variability. This provides a more actionable tool for the decision-making needs of adaptation planning, risk management, and early warning systems.

The remainder of the paper is organised as follows. Section~\ref{meth} introduces the proposed Bayesian VAR with multiple equality and inequality constraints as outlined in \citet{Chanetal2025}. Next, Section~\ref{forecasts} presents the results of our empirical study, including real time conditional temperature forecasts, while imposing equality and inequality constraints on a variety of emissions that correspond to different SSP scenarios, as well as a counterfactual study. Finally, Section~\ref{conclusion} provides some concluding remarks.

\section{Methodology}\label{meth}

As mentioned earlier, the purpose of this paper is to compute accurate forecasts of temperatures and various environmental variables aligned with the IPCC projections, from an optimistic scenario to a more pessimistic perspective. The novelty in the approach of this climate ex-ante exercise is that the multivariate model allows us to compare unconditional forecasts to conditional forecasts of specific variables of interest projected on the future paths of some other particular forcing variables.

\subsection{General Setup}
We briefly outline the approach of \cite{Chanetal2025} to produce unconditional and conditional forecasts, closely following their notation (see paper for further details). Consider first an $n \times 1$ vector of variables $\mathbf{y}_{t}=\left(y_{1, t}, \ldots, y_{n, t}\right)^{\prime}$ with a history $\mathbf{y}^{T}=\left(\mathbf{y}_{1-p}^{\prime}, \ldots, \mathbf{y}_{T}^{\prime}\right)^{\prime}$, and the $p$-lag (S)VAR:

\begin{equation}
\mathbf{A}_{0} \mathbf{y}_{t}=\mathbf{a}+\mathbf{A}_{1} \mathbf{y}_{t-1}+\cdots+\mathbf{A}_{p} \mathbf{y}_{t-p}+\varepsilon_{t}, \quad \varepsilon_{t} \sim \mathcal{N}\left(\mathbf{0}_{n}, \mathbf{I}_{n}\right) 
\end{equation}
with $\mathbf{a}$ an $n \times 1$ vector of intercepts, while $\mathbf{A}_{1}, \ldots,\mathbf{A}_{p}$ are the $n \times n$ VAR coefficient matrices and $\mathbf{A}_{0}$ a contemporaneous impact matrix.

Unconditional $h$-step ahead forecasts, $\mathbf{y}_{T+1, T+h}=\left(\mathbf{y}_{T+1}^{\prime}, \ldots, \mathbf{y}_{T+h}^{\prime}\right)^{\prime}$, are written as
\begin{equation}\label{unc}
\mathbf{H y}_{T+1, T+h}=\mathbf{c}+\varepsilon_{T+1, T+h}, \quad \varepsilon_{T+1, T+h} \sim \mathcal{N}\left(\mathbf{0}_{n h}, \mathbf{I}_{n h}\right) 
\end{equation}
with
\[
\mathbf{c}=\left[\begin{array}{c}
\mathbf{a}+\sum_{j=1}^{p} \mathbf{A}_{j} \mathbf{y}_{T+1-j}  \\
\mathbf{a}+\sum_{j=2}^{p} \mathbf{A}_{j} \mathbf{y}_{T+2-j} \\
\mathbf{a}+\sum_{j=3}^{p} \mathbf{A}_{j} \mathbf{y}_{T+3-j} \\
\vdots \\
\mathbf{a}+\mathbf{A}_{p} \mathbf{y}_{T} \\
\mathbf{a} \\
\vdots \\
\mathbf{a}
\end{array}\right], \mathbf{H}=\left[\begin{array}{cccccccc}
\mathbf{A}_{0} & \mathbf{0}_{n \times n} & \cdots & \cdots & \cdots & \cdots & \cdots & \mathbf{0}_{n \times n} \\
-\mathbf{A}_{1} & \mathbf{A}_{0} & \mathbf{0}_{n \times n} & \cdots & \cdots & \cdots & \cdots & \mathbf{0}_{n \times n} \\
-\mathbf{A}_{2} & -\mathbf{A}_{1} & \mathbf{A}_{0} & \mathbf{0}_{n \times n} & \cdots & & & \mathbf{0}_{n \times n} \\
\vdots & \ddots & \ddots & \ddots & \ddots & \ddots & & \vdots \\
-\mathbf{A}_{p-1} & \cdots & & -\mathbf{A}_{1} & \mathbf{A}_{0} & \mathbf{0}_{n \times n} & & \vdots \\
\mathbf{0}_{n \times n} & & & & \ddots & \ddots & \ddots & \vdots \\
\vdots & & \ddots & & \ddots & \ddots & \ddots & \vdots \\
\mathbf{0}_{n \times n} & \cdots & \mathbf{0}_{n \times n} & -\mathbf{A}_{p} & \cdots & -\mathbf{A}_{2} & -\mathbf{A}_{1} & \mathbf{A}_{0}
\end{array}\right]
\]
such that
\begin{equation}
\mathbf{y}_{T+1, T+h} \sim \mathcal{N}\left(\mathbf{H}^{-1} \mathbf{c},\left(\mathbf{H}^{\prime} \mathbf{H}\right)^{-1}\right) 
\end{equation}
Given that $\mathbf{H}$ is an $n h \times n h$ band matrix with band width $n p$, this makes the precision-based sampling approach of \cite{ChanJeliazkov2009} particularly convenient.

To construct conditional forecasts, we write these as a set of linear restrictions on the variables' future path:
\begin{equation}\label{rest}
\mathbf{R} \mathbf{y}_{T+1, T+h} \sim \mathcal{N}(\mathbf{r}, \boldsymbol{\Omega}) 
\end{equation}
such that $\mathbf{R}$ is a $r \times n h$ constant matrix with full row rank (ensuring there are no redundant restrictions), with $\mathbf{r}$ and $\boldsymbol{\Omega}$ representing the mean and covariance of the restrictions. Combining \eqref{unc} and \eqref{rest}, we get 
\begin{equation}\label{cond}
\mathbf{R} \mathbf{y}_{T+1, T+h}=\mathbf{R} \mathbf{H}^{-1} \mathbf{c}+\mathbf{R} \mathbf{H}^{-1} \varepsilon_{T+1, T+h} \sim \mathcal{N}(\mathbf{r}, \boldsymbol{\Omega}) 
\end{equation}

In order to derive restrictions on the future shocks implied by \eqref{rest} and \eqref{cond} as in \cite{Antolinetal2021}, let $\varepsilon_{T+1, T+h} \mid \mathbf{R}, \mathbf{r}, \boldsymbol{\Omega}$ denote the restricted future shocks with the distribution
\begin{equation}\label{dist}
\varepsilon_{T+1, T+h} \mid \mathbf{R}, \mathbf{r}, \boldsymbol{\Omega} \sim \mathcal{N}\left(\boldsymbol{\mu}_{\varepsilon}, \mathbf{I}_{n h}+\boldsymbol{\Psi}_{\varepsilon}\right), 
\end{equation}
with $\boldsymbol{\mu}_{\varepsilon}$ and $\boldsymbol{\Psi}_{\varepsilon}$ representing the deviations of the mean vector and covariance matrix of the \textit{restricted} future shocks from their \textit{unconditional} counterparts in \eqref{unc}. The above implies the restrictions on $\boldsymbol{\mu}_{\varepsilon}$ and $\boldsymbol{\Psi}_{\varepsilon}$ :
\begin{align}
\mathbf{R H}^{-1}\left(\mathbf{c}+\boldsymbol{\mu}_{\varepsilon}\right) & =\mathbf{r} \\
\mathbf{R H}^{-1}\left(\mathbf{I}_{n h}+\boldsymbol{\Psi}_{\varepsilon}\right) \mathbf{H}^{-1 \prime} \mathbf{R}^{\prime} & =\boldsymbol{\Omega}
\end{align}
with solution
\begin{align}
\boldsymbol{\mu}_{\varepsilon} & =\left(\mathbf{R} \mathbf{H}^{-1}\right)^{+}\left(\mathbf{r}-\mathbf{R} \mathbf{H}^{-1} \mathbf{c}\right) \notag \\ 
\boldsymbol{\Psi}_{\varepsilon} & =\left(\mathbf{R} \mathbf{H}^{-1}\right)^{+}\left(\boldsymbol{\Omega}-\mathbf{R}\left(\mathbf{H}^{\prime} \mathbf{H}\right)^{-1} \mathbf{R}^{\prime}\right)\left(\mathbf{R} \mathbf{H}^{-1}\right)^{+\prime} \tag{9}
\end{align}
where $\left(\mathbf{R H}^{-1}\right)^{+}$ is the Moore-Penrose inverse of $\mathbf{R} H^{-1}$. We should note that this solution minimizes the sum of the Frobenius norms of $\boldsymbol{\mu}_{\varepsilon}$ and $\boldsymbol{\Psi}_{\varepsilon}$, i.e. it returns the smallest deviations of the mean vector and covariance matrix between conditional and unconditional future shocks. Mapping the constraints on the shocks to the corresponding constraints on the forecasts, we have
\begin{align}
& \boldsymbol{\mu}_{\mathbf{y}}=\mathbf{H}^{-1}\left[\mathbf{c}+\left(\mathbf{R} \mathbf{H}^{-1}\right)^{+}\left(\mathbf{r}-\mathbf{R} \mathbf{H}^{-1} \mathbf{c}\right)\right]  \\
& \boldsymbol{\Sigma}_{\mathbf{y}}=\mathbf{H}^{-1}\left[\mathbf{I}_{n h}+\left(\mathbf{R} \mathbf{H}^{-1}\right)^{+}\left(\mathbf{\Omega}-\mathbf{R}\left(\mathbf{H}^{\prime} \mathbf{H}\right)^{-1} \mathbf{R}^{\prime}\right)\left(\mathbf{R} \mathbf{H}^{-1}\right)^{+\prime}\right] \mathbf{H}^{-1 \prime} .
\end{align}

In applications like ours, there is substantial uncertainty regarding the future path of some drivers of climate change we wish to condition our forecasts on. In these cases, this setup allows us to set the future values of the conditioned variables to lie within a certain range via inequality constraints:
\begin{equation}
\underline{\mathbf{c}}<\mathbf{S y}_{T+1, T+h}<\overline{\mathbf{c}}
\end{equation}
with $\mathbf{S}$ a $s \times n h$ pre-specified full-rank constant matrix, while $\underline{\mathbf{c}}$ and $\overline{\mathbf{c}}$ are $s \times 1$ vectors of constants, so that $\mathbf{y}_{T+1, T+h}$ has a truncated multivariate normal distribution
\begin{equation}\label{trunc}
\mathbf{y}_{T+1, T+h} \mid \underline{\mathbf{c}}<\mathbf{S y}_{T+1, T+h}<\overline{\mathbf{c}} \sim \mathcal{N}\left(\mathbf{H}^{-1} \mathbf{c},\left(\mathbf{H}^{\prime} \mathbf{H}\right)^{-1}\right) \mathbb{I}\left(\underline{\mathbf{c}}<\mathbf{S y}_{T+1, T+h}<\overline{\mathbf{c}}\right),
\end{equation}
where $\mathbb{I}$$(\cdot)$ is the indicator function.

\subsection{Conditional Forecasting: Constraints and Scenario Analysis}
To construct conditional forecasts of temperatures  given the future path of a subset of greenhouse gases' emissions, one can consider the case of conditional forecasts under equality constraints, as discussed in \citet{waggoner1999conditional}. This is represented as
\begin{equation}\label{rest2}
\mathbf{R}_{o} \mathbf{y}_{T+1, T+h}=\mathbf{r}_{o}
\end{equation}
where each row of $\mathbf{R}_{o}$ contains exactly one element that is 1 and all other elements are 0, while $\mathbf{r}_{o}$ is a  vector of constants, such that $\boldsymbol{\Omega}=\mathbf{0}_{r_{o} \times r_{o}}$. Here, the efficient sampling approach of \cite{chan2023high} together with the precision-based sampling approach of \cite{ChanJeliazkov2009} should be employed (see \citealp{Chanetal2025} for details).

So far we assumed that the restrictions are generated by all the structural shocks of the model, but this assumption could be relaxed and allow for the case in which we are interested in forecasts generated by restricting the path of a subset of structural shocks over the forecast horizon (see \citealp{BaumeisterKilian2014} and \citealp{Antolinetal2021}). This type of restriction can be formulated as
\begin{equation}\label{rest3}
\mathbf{W} \varepsilon_{T+1, T+h} \sim \mathcal{N}(\mathbf{w}, \mathbf{\Psi})
\end{equation}
where $\mathbf{W}$ is a full-rank selection matrix, $\mathbf{w}$ is a vector of constants and $\boldsymbol{\Psi}$ is a covariance matrix. 

Regarding (structural) scenario analysis, it combines constraints on future observations with the condition that only a subset of structural shocks deviate from their unconditional distribution, while the rest remain unchanged. This approach is more flexible and realistic than conditioning on a specific future path of structural shocks, which are unobserved and difficult to predict. It is also preferable to restricting only the future path of observables, as it allows users to specify which structural shocks drive future outcomes. Thus, combining restrictions on observables and restrictions on structural shocks, \eqref{rest2} and \eqref{rest3} allow us to restrict the path of future observables, so that these shocks retain their unconditional distribution: $\mathbf{W} \varepsilon_{T+1, T+h} \sim \mathcal{N}\left(\mathbf{0}_{w}, \mathbf{I}_{w}\right)$,which implies
\begin{equation}\label{dist2}
\mathbf{W H y}_{T+1, T+h} \sim \mathcal{N}\left(\mathbf{W} \mathbf{c}, \mathbf{I}_{w}\right).
\end{equation}
Combining \eqref{rest2} with \eqref{dist2} gives
\[
\underbrace{\left[\begin{array}{c}
\mathbf{R}_{o}  \tag{20}\\
\mathbf{W H}
\end{array}\right]}_{\widetilde{\mathbf{R}}} \mathbf{y}_{T+1, T+h} \sim \mathcal{N}(\underbrace{\left[\begin{array}{c}
\mathbf{r}_{o} \\
\mathbf{W} \mathbf{c}
\end{array}\right]}_{\widetilde{\mathbf{r}}}, \underbrace{\left[\begin{array}{cc}
\boldsymbol{\Omega}_{o} & \mathbf{0}_{r_{0} \times w} \\
\mathbf{0}_{w \times r_{0}} & \mathbf{I}_{w}
\end{array}\right]}_{\widetilde{\mathbf{\Omega}}}) .
\]
It can be seen that this case can be nested within the general framework in \eqref{rest} by setting $\mathbf{R}=\widetilde{\mathbf{R}}$, $\mathbf{r}=\widetilde{\mathbf{r}}$ and $\boldsymbol{\Omega}=\widetilde{\boldsymbol{\Omega}}$.

\section{Conditional Forecasts and Scenario Analysis for Global Temperatures}\label{forecasts}
We now apply the methodology discussed above, employing a Bayesian VAR with an asymmetric conjugate prior \citep{chan2022asymmetric} and $n=8$ annual variables aiming to examine how key climate variables dynamically evolve over time.\footnote{The asymmetric conjugate prior was chosen as it can accommodate cross‐variable shrinkage, while being able to maintain analytical results, like the closed‐form expression of the marginal likelihood. Results remain robust under alternative priors, including a Minnesota type prior, as that can be seen in Figures~\ref{Figure: Adverse Scenario MP} $\&$~\ref{Figure: Optimistic Scenario MP} in the Appendix.} In particular, we are interested in examining the evolution of temperature anomalies and greenhouse gases (both natural and anthropogenic), while at the same time controlling for solar irradiance and natural aerosols. The complete list of variables that are used in the model can be seen in Table~\ref{Table: Data}. Our choice of variables reflects the need to strike a balance between incorporating temperatures plus their main drivers (see \citealp{agliardi_relationship_2019} and \citealp{Phellaetal2024}, for example) while keeping the dimension of the VAR manageable.\footnote{One possibility would be to have \textit{Solar} as an exogenous variable, i.e. outside the VAR, which would help further reduce the dimension of the VAR. However, having this variable in the VAR seems to help in terms of the model's forecasting ability.}

The data comes mostly from \cite{Meinshausenetal2020} -- in their work, these authors provide historical annual averages for the relevant variables up to 2014, then climate models-based projections from 2015 onwards.\footnote{See \href{https://www.climatecollege.unimelb.edu.au/cmip6}{https://www.climatecollege.unimelb.edu.au/cmip6}.} These are in accordance with the different Shared Socioeconomic Pathways (SSPs) scenarios we utilise to set the future path for several emissions, and are available from 2015 to 2500, although our forecast horizon is until 2050. Nevertheless, we extend the sampling period of \textit{actual} realisations to 2023 using data from NOAA, such that our sample spans from 1850 up until 2023.

\begin{table}[h]
\caption{Climate variables}
\label{Table: Data}
\resizebox{0.99\textwidth}{!}{
\begin{tabular}{llll}
\hline \hline
\multicolumn{1}{c}{\textbf{Variable}} & \multicolumn{1}{c}{\textbf{Full Description}}           & {\textbf{Unit}} & \textbf{Source}    \\ \hline

$Solar$ & Solar Irradiance & No. of sunspots &  Royal Observatory of Belgium  \\
$Temp$   & Global temperature anomalies    &  $^\circ C$    & \cite{Meinshausenetal2020}, NOAA \\
$WMGHG$ & Well-mixed greenhouse gases  & $W/m2$ & \cite{Meinshausenetal2020}, NOAA  \\

$AN$ & Aero Naturals & $W/m2$ & NOAA \\ 
$AS$ & Aerosols & $W/m2$ & NOAA\\ 
$CO_2$ & Carbon dioxide emissions & ppm & \cite{Meinshausenetal2020}, NOAA  \\ 
$CH_4$ & Methane emissions  &ppb & \cite{Meinshausenetal2020}, NOAA  \\ 
$N_2O$ & Nitrous Oxide & ppb & \cite{Meinshausenetal2020}, NOAA \\ \hline \hline
\end{tabular}
}
\justify
\flushleft{
\footnotesize{Notes: $^\circ C$ denotes degrees Celsius, ppp is parts per million, ppb is parts per billion, $W/m^2$ is watts per square metre, NOAA is the National Oceanic and Atmospheric Administration.}}%
  \end{table}

\begin{figure}[h]
\centering
\hspace*{-2.5cm}  
\includegraphics[width=1.25\textwidth]{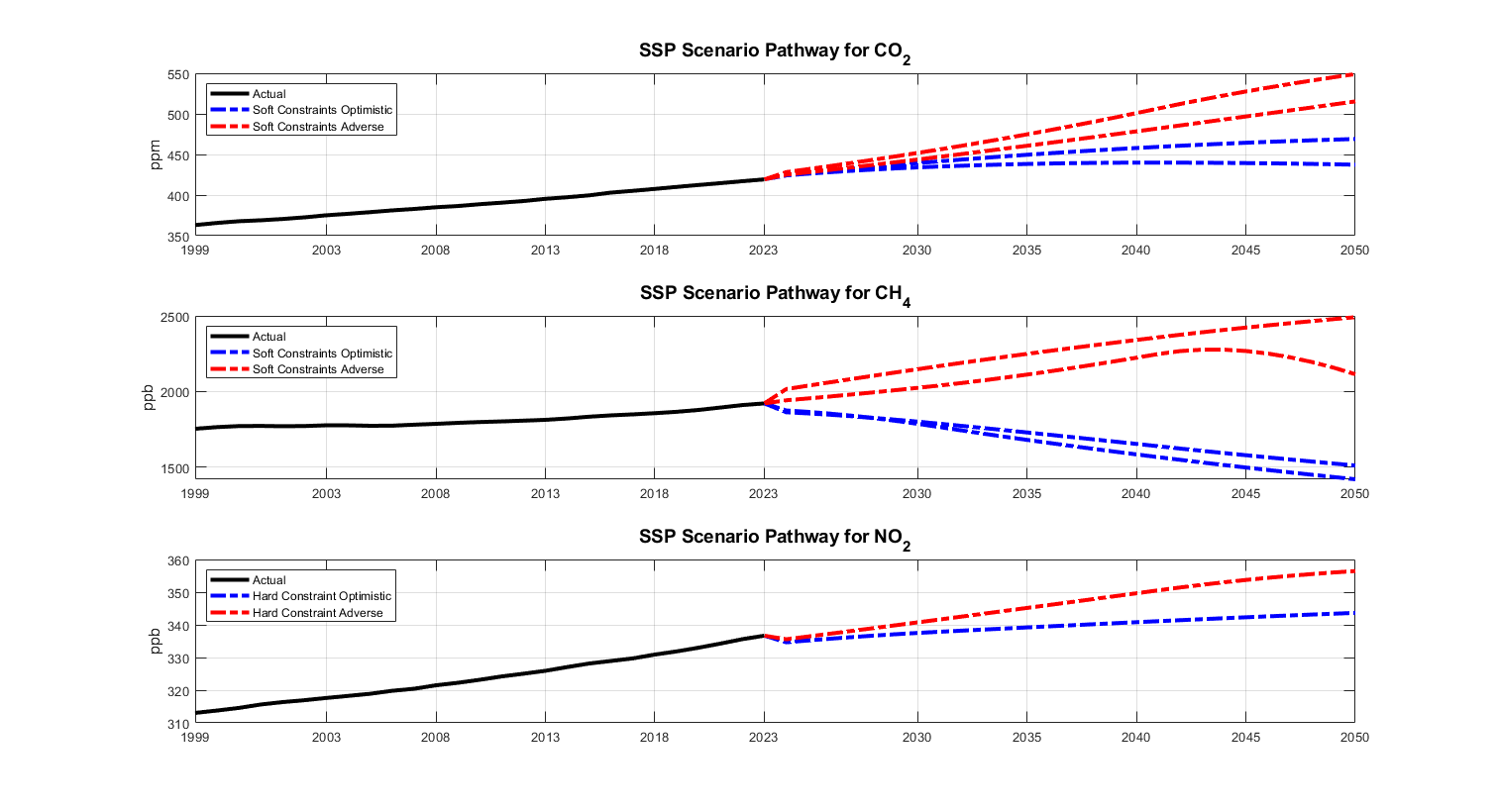}
\caption{\textit{Equality and inequality constrains for $CO_2$, $CH_4$ and $N_2O$ under an adverse and an optimistic scenario up until 2050.}}
\label{Figure: Scenario Pathways}
\end{figure}

In practice, we consider two different forecasting scenarios, namely (i) \textit{adverse}, and (ii) \textit{optimistic}, where an inequality constraint is imposed on the future path of carbon dioxide ($CO_2$) and methane ($CH_4$), while strict equality constraints are imposed on the path of nitrus oxide ($N_2O$).\footnote{We impose the inequality constraints on $CO_2$ emissions given that it forms the bulk of greenhouse gases' emissions and is usually the focus of policy interventions. Meanwhile, though methane has a shorter atmospheric lifespan than $CO_2$, its warming effect is over 80 times stronger on a per-unit mass basis over a 20-year period, and thus also crucial for climate policy \citep{globalmethanepledge2025}. The dataset in \cite{Meinshausenetal2020} allows for a much more comprehensive study of climate change drivers.}
The adverse scenario conditions the future paths of $CO_2$, $CH_4$ and $N_2O$ to the corresponding values from SSP scenarios that imply  a world focused on economic growth and technological advancement at the expense of environmental sustainability, therefore with little to no mitigation and high emissions (i.e., SSP 4-6 $\&$ SSP 5-8.5), while the optimistic scenario conditions on SSP scenario values that imply ambitious mitigation strategies and achieving lower emission targets, in line with the Paris Agreement (i.e., SSP 1-1.9 $\&$ SSP 1-2.6). Figure ~\ref{Figure: Scenario Pathways} summarises the inequality and equality constraints for both the optimistic and adverse scenario, while the exact values can also be seen in Tables~\ref{Table: Scenario Values Adverse} $\&$~\ref{Table: Scenario Values Optimistic} in the Appendix.
\bigskip

\subsection{Forecasting Performance}
Before engaging in a real-time forecasting exercise, given the availability of SSP scenarios from 2015 onward, we conduct a preliminary pseudo-out-of-sample forecasting exercise to compare our chosen approach with alternative models. It is important to note that our framework does not lend itself to a direct comparison with most existing forecasting approaches. On the one hand, temperature forecasts from climate models (as produced by meteorological offices) can incorporate emissions scenarios but differ fundamentally from our `reduced-form' methodology. On the other hand, standard reduced-form models (e.g., AR or ARDL specifications) cannot easily accommodate future scenario values that involve inequality constraints on emissions.

Despite these caveats, it is useful to assess how our model fares against alternative approaches. For simplicity, we impose a ‘business-as-usual’ (i.e., adverse) scenario for our conditioning variables from 2016–2023 and compare the resulting one-year-ahead forecasts against simple reduced-form alternatives (i.e., AR(4), ARDL) and those from a suite of physical models compiled by the World Meteorological Organization (WMO) Lead Lead Centre for Annual-to-Decadal Climate Prediction, hosted by the UK Met Office.\footnote{This centre produces a consolidated, multi-model forecast, integrating predictions from four designated \textit{Global Producing Centres}---the Met Office (UK), Barcelona Supercomputing Centre (BSC, Spain), the Canadian Centre for Climate Modelling and Analysis (CCCma, Canada), and the Deutscher Wetterdienst (DWD, Germany)---alongside contributions from around 15 other forecast groups worldwide, each running dynamical climate models, with multiple ensemble members (e.g., 190--220 models in recent years) to capture a range of possible outcomes.} We choose to impose a ‘business-as-usual’ (i.e., adverse) scenario for emissions in our conditional forecasts, as this provides the closest analogue to the reduced-form benchmark models (such as AR or ARDL), which condition directly on the actual realizations of the variables as they become available. By adopting this scenario, our conditional forecasts remain comparable to those benchmarks while also allowing us to highlight the additional advantage of our approach—namely, the ability to incorporate uncertainty in future emissions trajectories.

The full set of forecast performance measures is reported in Table~\ref{Table: Forecast Performance Measures}. Two main results emerge. First, our forecasts perform at least as well as standard reduced-form alternatives. More importantly, our methodology outperforms a range of physical models from the WMO, while additionally offering the advantage of incorporating uncertainty in future emissions.\footnote{The short span of available SSP scenarios up to 2023, the point at which our real-time forecasting exercise begins, limits the evaluation window and does not permit statistically powerful tests of relative forecasting performance.} These results suggest that our model offers a reasonable alternative to existing temperature forecasting models.

\begin{table}[h]
\caption{Forecast Performance Measures}
\label{Table: Forecast Performance Measures}
\center
\small{
\begin{tabular}{lccccccc}
\hline \hline
\textbf{Model} & \multicolumn{2}{c}{\textbf{Unconditional}} & \multicolumn{2}{c}{\textbf{\begin{tabular}[c]{@{}c@{}}Conditional \\ (business as usual)\end{tabular}}} & \multicolumn{2}{c}{\textbf{AR(4)}} & \textbf{ARDL}        \\
\textbf{MSE}   & \multicolumn{2}{c}{0.0211}                 & \multicolumn{2}{c}{0.0244}                                                                              & \multicolumn{2}{c}{0.0201}         & 0.0191               \\
\textbf{MAE}   & \multicolumn{2}{c}{0.1172}                 & \multicolumn{2}{c}{0.1314}                                                                              & \multicolumn{2}{c}{0.1148}         & 0.1062               \\
\textbf{}      &                     &                      &                                                     &                                                   &                  &                 &                      \\
\textbf{Model} & \textbf{BSC}        & \textbf{CCCma}        & \textbf{DWD/MPI}                                    & \textbf{MIROC}                                    & \textbf{MOHC}    & \textbf{MRI}    & \textbf{Multi-model} \\
\textbf{MSE}   & 0.0454              & 0.0402               & 0.0512                                              & 0.0314                                            & 0.0286           & 0.0438          & 0.0301               \\
\textbf{MAE}   & 0.1784              & 0.1693               & 0.1950                                              & 0.1313                                            & 0.1324           & 0.1650          & 0.1293   \\           \hline \hline
\end{tabular}
}
\justify
\flushleft{
\footnotesize{Notes: AR(4) denotes an autoregressive model of order 4; ARDL is an autoregressive distributed lag model incorporating the variables in Table 1 as regressors (lag orders have been chosen under standard information criteria); BSC is the Barcelona Computing Centre, CCCma is the Canadian Centre for Climate Modelling and Analysis, DWD is the Deutscher Wetterdienst, MPI is the Max Planck Institute, MIROC is the Model for Interdisciplinary Research on Climate, MOHC is the Met Office Hadley Centre, MRI is the Metereological Research Institute, `Multi-model' denotes the multi-model ensemble mean computed by the World Metereological Organization Lead Centre for Annual to Decadal Climate Prediction.}}%
\end{table}

\subsection{Real-time Temperature Forecasts}
Building on the pseudo-out-of-sample evaluation, we next turn to a real-time forecasting exercise, where in this instance we will be considering both the adverse and optimistic forecasting scenarios. In both cases we estimate our model from 1850 to 2023 and then impose the corresponding paths on the three emission variables, while examining the dynamic evolution of temperatures and well-mixed greenhouse gases for the duration of the forecasting period.\footnote{All climate variables presented in Table~\ref{Table: Data} are included in the model, though the main focus is on the two selected variables we present, namely temperatures and well-mixed greenhouse gases.} Taking into consideration that the relationship between these variables evolves slowly, but cautious of the curse of dimensionality in such a multivariate setup, we set the number of lags in the model equal to 4. Results, however, remain robust across different lag orders.

\begin{figure}[H]
\centering
\hspace*{-2.5cm}  
\includegraphics[width=1.25\textwidth]{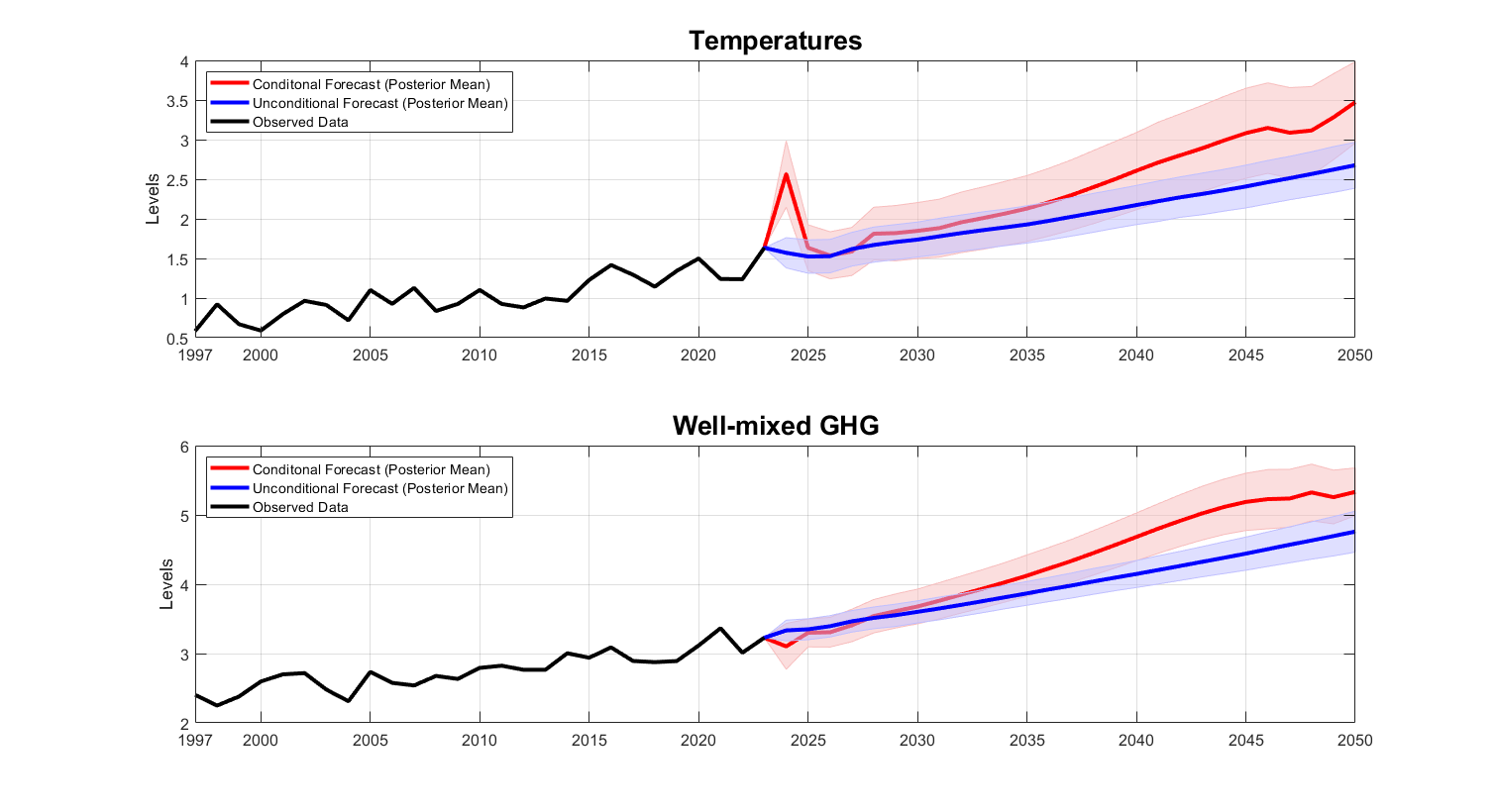}
\caption{\textit{Conditional and unconditional forecasts when $CO_2$, $CH_4$ and $N_2O$ emissions in 2024-2050 match the SSP adverse scenario projections. The shaded bands correspond to the $68\%$ coverage intervals while the solid black lines denote the in-sample values.}}
\label{Figure: Adverse Scenario}
\end{figure}

\begin{figure}[H]
\centering
\hspace*{-2.5cm}  
\includegraphics[width=1.25\textwidth]{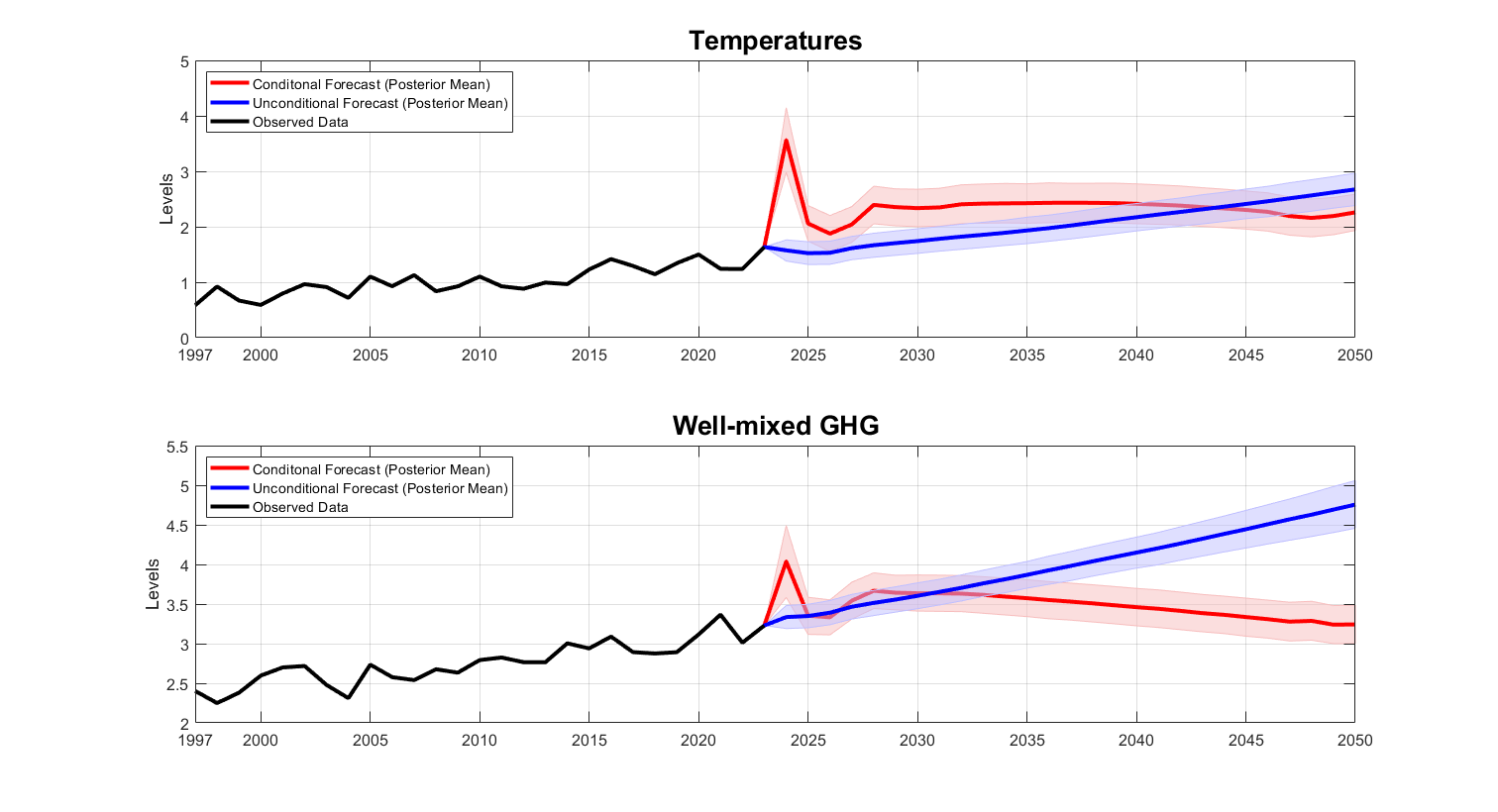}
\caption{\textit{Conditional and unconditional forecasts when $CO_2$, $CH_4$ and $N_2O$ emissions in 2024-2050 match the SSP optimistic scenario projections. The shaded bands correspond to the $68\%$ coverage intervals while the solid black lines denote the in-sample values.}}
\label{Figure: Optimistic Scenario}
\end{figure}

Figures~\ref{Figure: Adverse Scenario} and~\ref{Figure: Optimistic Scenario} display the actual realisations up to 2023 (black line), together with  the unconditional forecasts (blue solid line and bands) and conditional forecasts (red solid line and bands) for the variables of interest up until 2050. The bands correspond to the $68\%$ coverage intervals, while the solid line corresponds to the posterior means.\footnote{In order to reconcile the jump between the last realisation of the conditioning variables and the first forecasting period where the SSP scenario values are imposed, the model may generate sharp jumps in the first period of the forecasting sample which should be disregarded. The reader is advised to rather focus on the dynamic evolution of $Temperatures$ and $WMGHG$.} 

\begin{figure}[h]
\centering
\hspace*{-2.5cm}  
\includegraphics[width=1.25\textwidth]{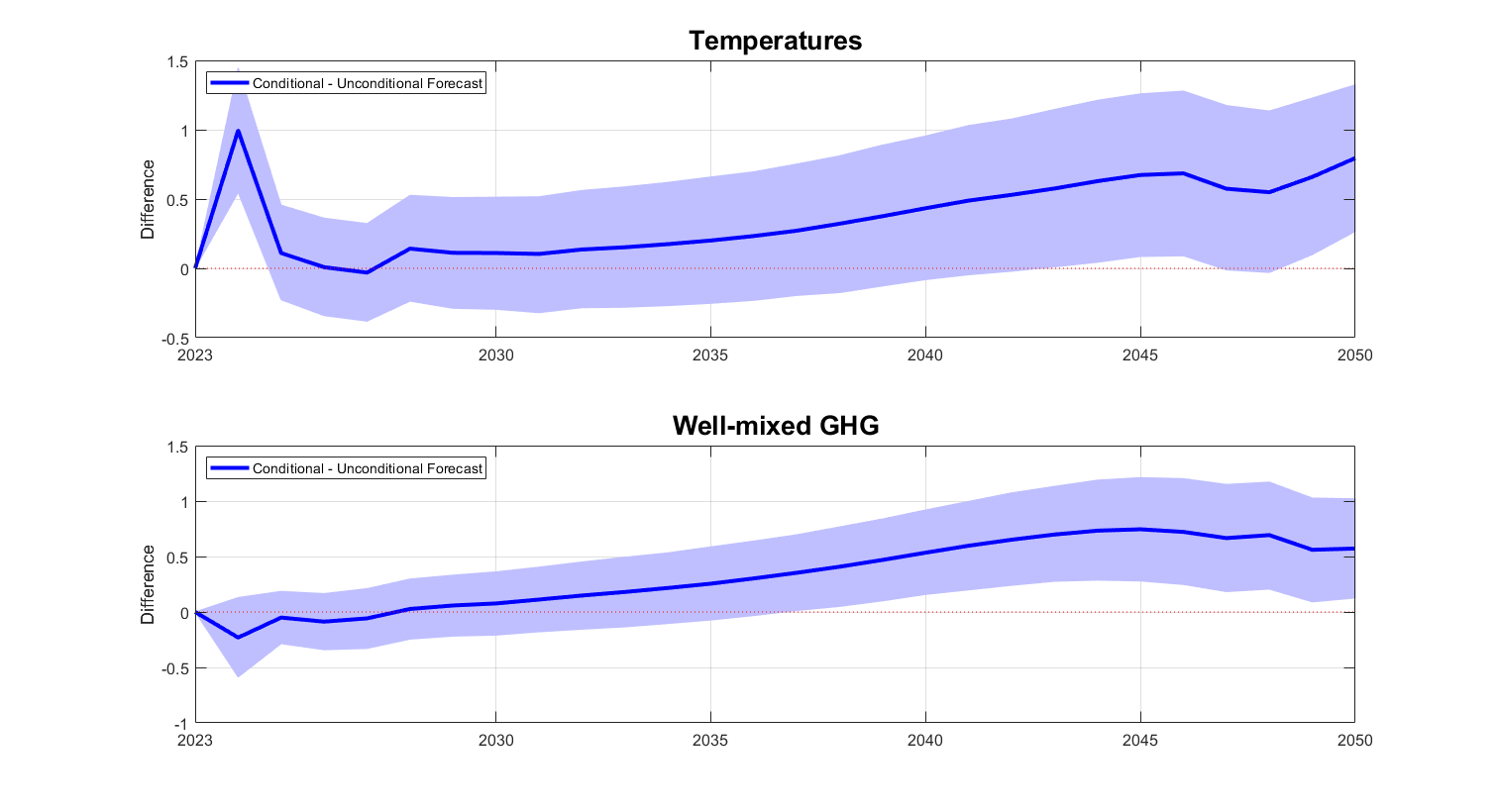}
\caption{\textit{Difference between the conditional and unconditional forecasts when $CO_2$, $CH_4$ and $N_2O$ emissions in 2024-2050 match the SSP adverse scenario projections. The shaded bands correspond to the $68\%$ coverage intervals.}}
\label{Figure: Adverse Scenario Difference}
\end{figure}

Figure~\ref{Figure: Adverse Scenario} displays the results for an adverse scenario where the world is focused on rapid economic growth, technological advancement, and high energy consumption, with heavy reliance on fossil fuels (coal, oil, gas) to power industries and transportation. While the world is \enquote{Taking-the-Highway}, sustainability is not a priority resulting to high emissions. As seen in the plots, in such a case the conditional forecasts resemble the trajectory of unconditional forecasts, which essentially imply the dynamic evolution of climate variables will follow a \enquote{business-as-usual} path. This can also be seen in Figure~\ref{Figure: Adverse Scenario Difference}, which plots the posterior differences between the conditional and unconditional forecasts, along with the posterior means (solid line) and distributional $68\%$ coverage intervals. As it can be seen in this case the two paths are not significantly different from each other. Under this scenario therefore, we would see temperature anomalies that reach close to  3$^\circ C$ by the year 2050, almost double the threshold target set by the Paris Agreement.

On the other hand, if the world turned into \enquote{Taking-the-Green-Road} scenario instead -- by following an optimistic scenario where ambitious climate mitigation strategies could be adopted with rapid global action to reduce emissions, transition to renewable energy, and implementation of sustainable policies -- this would imply a halt in the rise in temperature anomalies. Figure~\ref{Figure: Optimistic Scenario} demonstrates how, under such a scenario, conditional forecasts do not seem to follow a similar trajectory to the unconditional forecasts but rather remain more stable and, in practice, temperature anomalies could be kept below $2^\circ C$ and close to the Paris Agreement target of $1.5^\circ C$.\footnote{Conditional temperatures are higher than their unconditional counterparts at the beginning of the forecasting sample due to the effort of the model to reconcile the jump between the last realisation of emissions and the significantly lower value at which we condition in the first forecasting period, and as such the focus should remain on the fact that temperatures remain relatively stable. In practice, it is expected that at the beginning of the forecasting sample the conditional model would experience a slight increase in temperature anomalies, like in the unconditional case, that would then plateau.} However, even if the world managed to transition into this extremely ambitious scenario, the plateauing of temperature anomalies would take more than 20 years to occur. 

\begin{figure}[h]
\centering
\hspace*{-2.5cm}  
\includegraphics[width=1.25\textwidth]{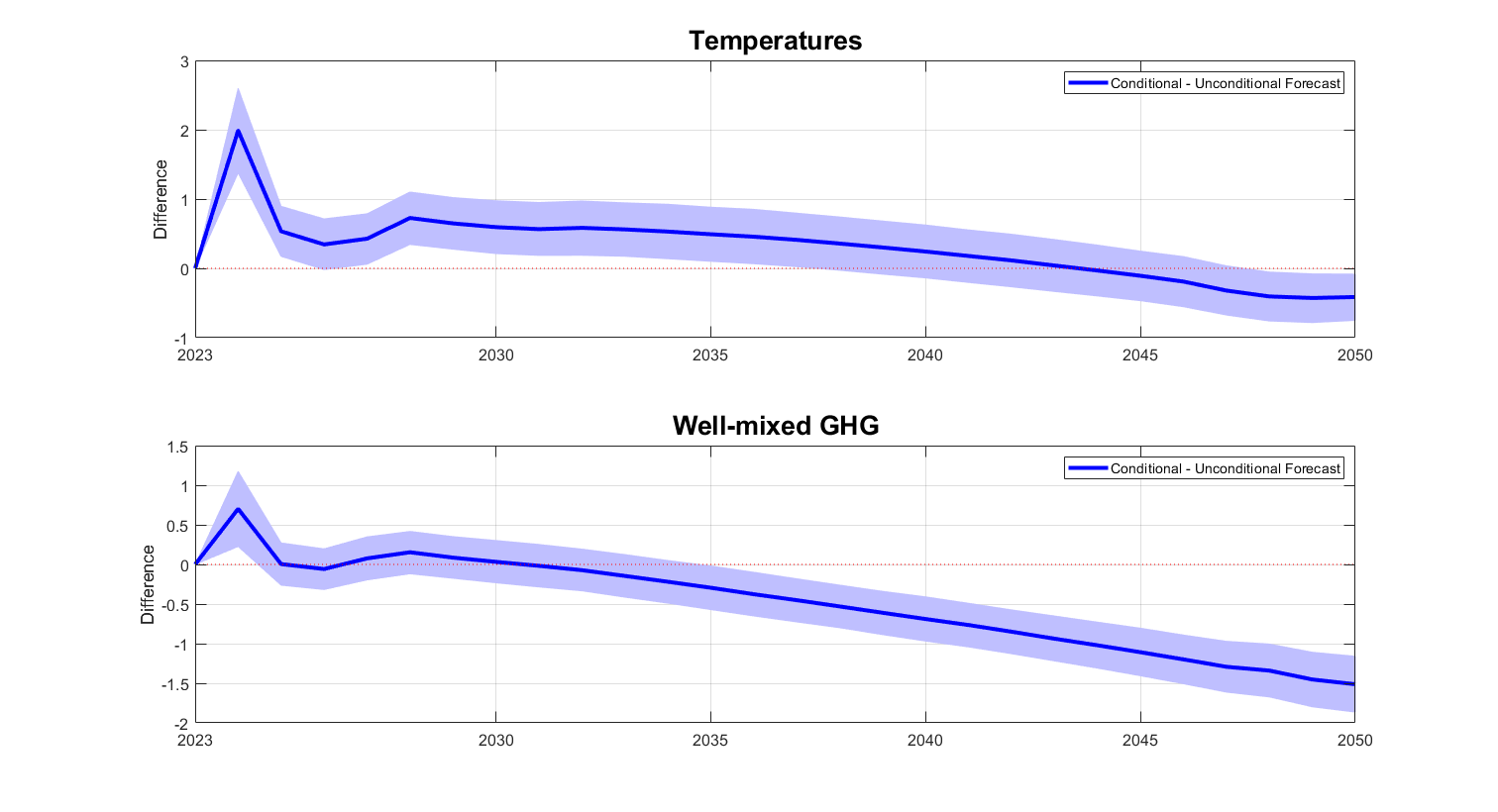}
\caption{\textit{Difference between the conditional and unconditional forecasts when $CO_2$, $CH_4$ and $N_2O$ emissions in 2024-2050 match the SSP optimistic scenario projections. The shaded bands correspond to the $68\%$ coverage intervals.}}
\label{Figure: Optimistic Scenario Difference}
\end{figure}
As it can be seen in Figure~\ref{Figure: Optimistic Scenario Difference}, which plots the differences between the conditional and unconditional forecasts under this scenario, the two paths will be significantly different from each other around 2047. Another striking feature is the prediction that well-mixed greenhouse gases would be, under such conditions, significantly lower well before 2035, compared to their unconditional counterparts. This could imply additional gains further in the future that cannot be captured within the time period examined, given the long run relationship between these climate variables.

\begin{figure}[h]
\centering
\hspace*{-2.5cm}  
\includegraphics[width=1.25\textwidth]{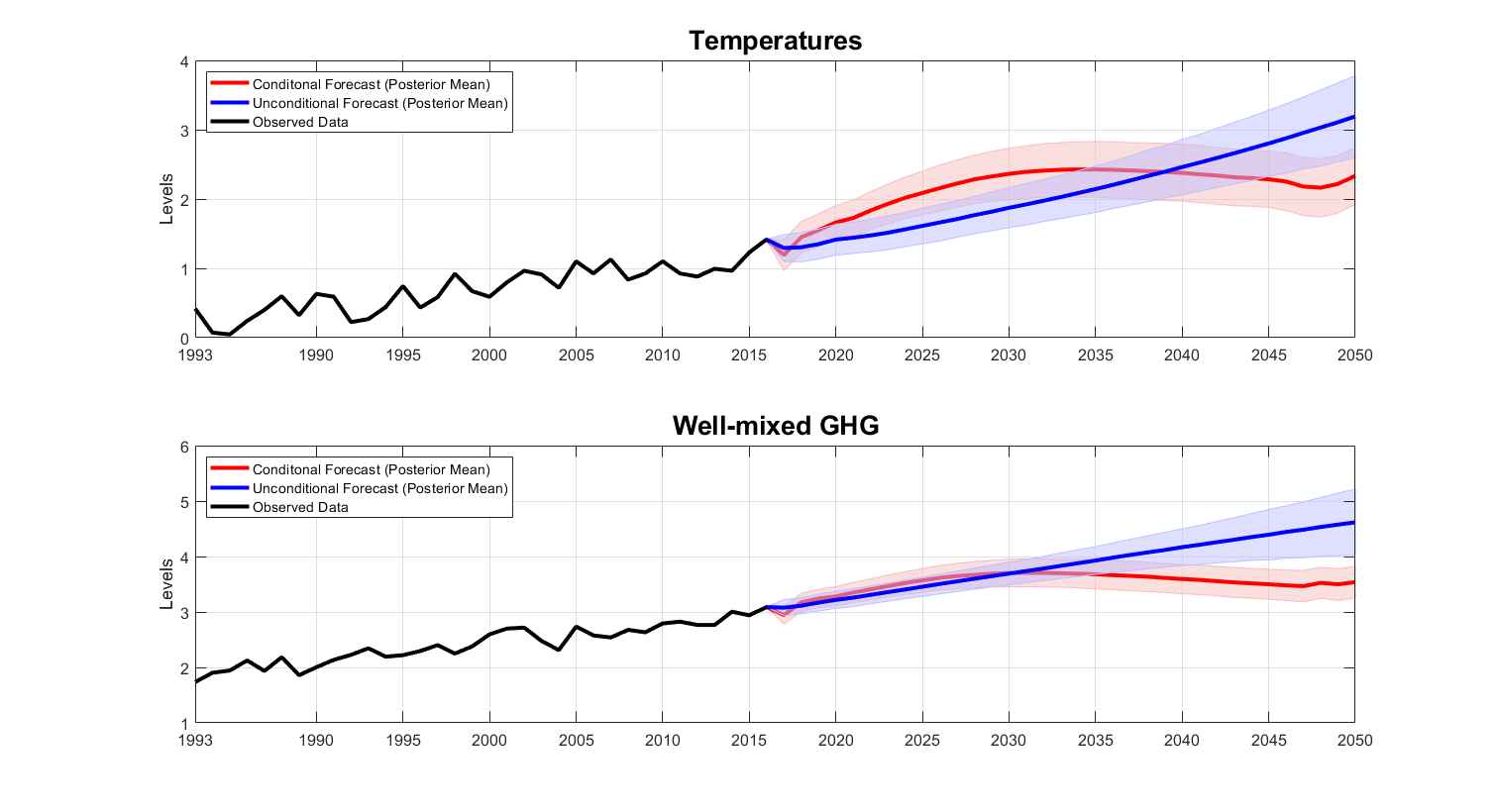}
\caption{\textit{Conditional and unconditional forecasts when $CO_2$, $CH_4$ and $N_2O$ emissions in 2016-2050 match the SSP optimistic scenario projections. The shaded bands correspond to the $68\%$ coverage intervals while the solid black lines denote the in-sample values.}}
\label{Figure: Counterfactual Optimistic Scenario}
\end{figure}

\begin{figure}[h]
\centering
\hspace*{-2.5cm}  
\includegraphics[width=1.25\textwidth]{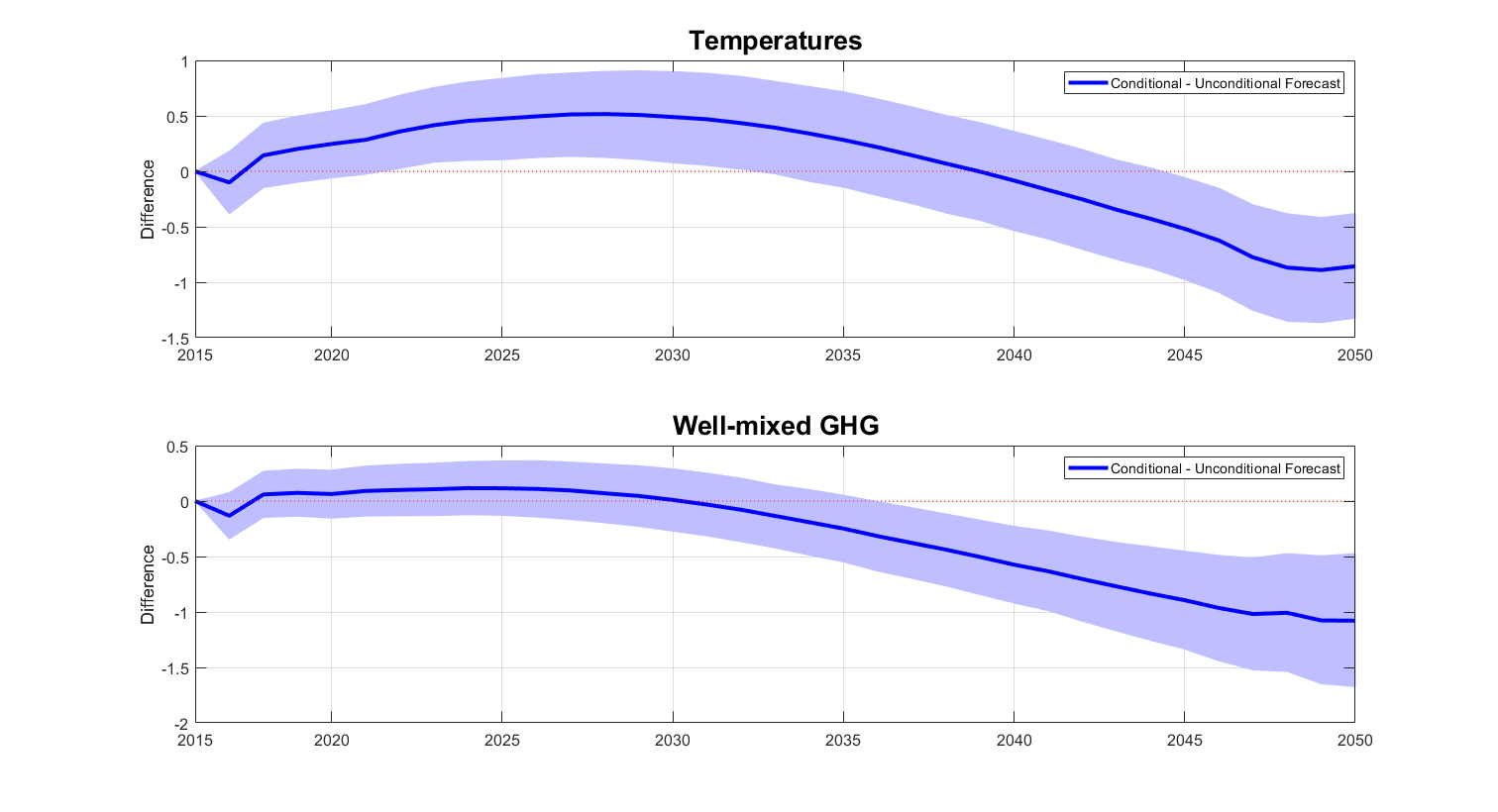}
\caption{\textit{Difference between the conditional and unconditional forecasts when $CO_2$, $CH_4$ and $N_2O$ emissions in 2016-2050 match the SSP optimistic scenario projections. The shaded bands correspond to the $68\%$ coverage intervals.}}
\label{Figure: Counterfactual Optimistic Scenario Difference}
\end{figure}

Given the apparent gains from achieving conditions similar to those outlined under the optimistic scenario, we complement the real-time forecasting exercise with a counterfactual analysis that aims to examine how temperatures would have evolved if this optimistic scenario was in fact implemented earlier. Given the availability of SSP scenario emissions values following 2015, we produce a counterfactual forecasting exercise from 2016 onwards under the optimistic conditions outlined before. Figure~\ref{Figure: Counterfactual Optimistic Scenario} displays the evolution of the two key variables of interest under this counterfactual exercise. It can be clearly seen that if a high mitigation, low emissions, scenario had been adopted even just a few years ago, the stabilisation of temperature anomalies could have been brought forward by approximately 5 years (i.e., the crossing point between conditional and unconditional projections) and the drop in temperature levels would have been more noteworthy.\footnote{As one could note here, the initial jump in the projections is significantly smaller in this case. This is due to the fact that in the earlier years of the SSP scenarios, the realised emissions were more comparable to the conditional emission values under an optimistic scenario and, in such a case, at the beginning of the forecasting sample, the conditional and unconditional paths are expected to be rather similar to each other. Hence, the argument brought forward in the previous figures that the main focus should be on the long-run evolution of these climate variables.} This also becomes evident in Figure~\ref{Figure: Counterfactual Optimistic Scenario Difference}, as the difference across the two forecasts becomes significant from around 2043, instead of the previous year of 2048. Nevertheless, this result highlights that the benefits of stabilizing temperatures earlier do not scale proportionally with the duration of reduced emissions. This insight could provide a useful perspective for policymakers, highlighting that the timing and design of emission reduction targets may matter as much as their stringency which could help refine the balance between ambition and feasibility in policy design.

\section{Conclusion}\label{conclusion}
Forecasting temperatures under different SSP scenarios is crucial for understanding potential climate futures and guiding policy and adaptation efforts. As we attempt to illustrate here, VAR models provide a valuable tool for analyzing the interdependencies between climate and socioeconomic variables, offering insights into how different pathways may influence global temperature trends.

This paper produces real-time, ex-ante forecasts for key climate variables, namely temperature anomalies, by conditioning on a number of climate change drivers. In particular we provide forecasts up until 2050, by conditioning on the future path of emissions ($CO_2$, $CH_4$ and $N_2O$) as those are specified in different SSP scenarios. We explore the two extreme cases, one of a world with little to no mitigation with high emissions and a more optimistic alternative with ambitious mitigation policies and low emissions. The results show that in a \enquote{business-as-usual} world, the conditional forecasts produced follow to a large extend the trajectory of the unconditional forecasts, and predict a rise in temperature anomalies reaching almost $3^\circ C$, a value that is almost double the Paris Agreement target. On the contrary if the world was instead able to \enquote{take-the-green-road}, temperature anomalies would actually plateau below $2^\circ C$. Furthermore, if the world was able to adopt such an optimistic scenario earlier than today, the stabilisation of temperatures could in fact be achieved almost 5 years earlier. 

An alternative way to interpret our results is to think of these as forecast ranges for temperatures in a context of uncertainty about the path of key drivers. Another possibility, easily accommodated by this framework, is to study the predicted path of temperatures if policymakers were to cap emissions at a certain level (say, 2024 levels). In the same vein, we could be interested in restricting the future path of temperature anomalies to be, say, between $1.^\circ C$ and $2^\circ C$ for the next 10 years and between $1.5^\circ C$ and $1^\circ C$ afterwards, and back out the required level of emissions consistent with such a scenario. Within our general setup, imposing inequality constraints on observables can be seamlessly formulated using equation \eqref{trunc}. Specifically, this scenario can be implemented by setting $\mathbf{S} = \mathbf{I}_{n h}$ and appropriately selecting the relevant elements in $\underline{\mathbf{c}}$. We leave this for future research.

\newpage
\bibliographystyle{authordate1}
\bibliography{Bibliography}

\section{Appendix}

\begin{table}[H]
\caption{Summary of equality and inequality constraints under an adverse scenario}
\label{Table: Scenario Values Adverse}
\centering
\begin{tabular}{lcccc} \hline \hline
     & \multicolumn{2}{c}{{\underline{$CO_2$ Inequality Constraint}}}                                                            & \multicolumn{2}{c}{{\underline{Equality Constraint}}} \\ 
Date & \begin{tabular}[c]{@{}c@{}}$CO_2$ Lower Bound\\ (SSP4-6)\end{tabular}   & \begin{tabular}[c]{@{}c@{}}$CO_2$ Upper Bound\\ (SSP 5-8.5)\end{tabular} & \begin{tabular}[c]{@{}c@{}}$CH_4$\\ (SSP 5-8.5)\end{tabular}   & \begin{tabular}[c]{@{}c@{}}$N_2O$\\ (SSP 5-8.5)\end{tabular}   \\ \hline
2024 & 417.235                                                              & 428.297                                                               & 1942.492               & 335.587               \\
2025 & 419.189                                                              & 431.957                                                               & 1954.742               & 336.433               \\
2026 & 421.157                                                              & 435.727                                                               & 1967.639               & 337.285               \\
2027 & 423.132                                                              & 439.606                                                               & 1981.133               & 338.143               \\
2028 & 425.105                                                              & 443.593                                                               & 1995.174               & 339.006               \\
2029 & 427.072                                                              & 447.691                                                               & 2009.713               & 339.874               \\
2030 & 429.033                                                              & 451.897                                                               & 2024.709               & 340.747               \\
2031 & 430.989                                                              & 456.214                                                               & 2040.114               & 341.626               \\
2032 & 432.965                                                              & 460.654                                                               & 2056.384               & 342.509               \\
2033 & 434.990                                                              & 465.228                                                               & 2073.889               & 343.393               \\
2034 & 437.076                                                              & 469.934                                                               & 2092.547               & 344.280               \\
2035 & 439.228                                                              & 474.774                                                               & 2112.271               & 345.171               \\
2036 & 441.447                                                              & 479.745                                                               & 2132.959               & 346.063               \\
2037 & 443.729                                                              & 484.849                                                               & 2154.534               & 346.958               \\
2038 & 446.066                                                              & 490.089                                                               & 2176.914               & 347.855               \\
2039 & 448.445                                                              & 495.462                                                               & 2200.041               & 348.756               \\
2040 & 450.863                                                              & 500.972                                                               & 2223.834               & 349.658               \\
2041 & 453.316                                                              & 506.619                                                               & 2248.224               & 350.562               \\
2042 & 455.811                                                              & 512.206                                                               & 2267.658               & 351.441               \\
2043 & 458.349                                                              & 517.548                                                               & 2276.962               & 352.262               \\
2044 & 460.931                                                              & 522.665                                                               & 2276.892               & 353.027               \\
2045 & 463.574                                                              & 527.564                                                               & 2268.134               & 353.736               \\
2046 & 466.290                                                              & 532.255                                                               & 2251.316               & 354.389               \\
2047 & 469.079                                                              & 536.743                                                               & 2227.045               & 354.989               \\
2048 & 471.940                                                              & 541.034                                                               & 2195.857               & 355.533               \\
2049 & 474.866                                                              & 545.130                                                               & 2158.271               & 356.023               \\
2050 & 477.845                                                              & 549.033                                                               & 2114.744               & 356.460      \\ \hline \hline        
\end{tabular}
\end{table}

\begin{table}[H]
\caption{Summary of equality and inequality constraints under an optimistic scenario}
\label{Table: Scenario Values Optimistic}
\centering
\begin{tabular}{lcccc} \hline \hline
     & \multicolumn{2}{c}{{\underline{$CO_2$ Inequality Constraint}}}                                                            & \multicolumn{2}{c}{{\underline{Equality Constraint}}} \\ 
Date & \begin{tabular}[c]{@{}c@{}}$CO_2$ Lower Bound\\ (SSP1-1.9)\end{tabular}   & \begin{tabular}[c]{@{}c@{}}$CO_2$ Upper Bound\\ (SSP 1-2.6)\end{tabular} & \begin{tabular}[c]{@{}c@{}}$CH_4$\\ (SSP 1-1.9)\end{tabular}   & \begin{tabular}[c]{@{}c@{}}$N_2O$\\ (SSP 1-1.9)\end{tabular}   \\ \hline
2024 & 424.222                                                              & 424.899                                                               & 1875.310               & 334.653               \\
2025 & 426.303                                                              & 427.451                                                               & 1866.105               & 335.220               \\
2026 & 428.203                                                              & 429.942                                                               & 1854.492               & 335.750               \\
2027 & 429.927                                                              & 432.373                                                               & 1840.639               & 336.245               \\
2028 & 431.481                                                              & 434.748                                                               & 1824.707               & 336.705               \\
2029 & 432.871                                                              & 437.067                                                               & 1806.825               & 337.130               \\
2030 & 434.098                                                              & 439.335                                                               & 1787.121               & 337.520               \\
2031 & 435.167                                                              & 441.551                                                               & 1765.736               & 337.876               \\
2032 & 436.110                                                              & 443.698                                                               & 1743.814               & 338.214               \\
2033 & 436.951                                                              & 445.758                                                               & 1722.428               & 338.549               \\
2034 & 437.689                                                              & 447.736                                                               & 1701.541               & 338.880               \\
2035 & 438.325                                                              & 449.632                                                               & 1681.141               & 339.209               \\
2036 & 438.855                                                              & 451.450                                                               & 1661.179               & 339.534               \\
2037 & 439.284                                                              & 453.190                                                               & 1641.634               & 339.857               \\
2038 & 439.608                                                              & 454.855                                                               & 1622.478               & 340.177               \\
2039 & 439.830                                                              & 456.446                                                               & 1603.680               & 340.495               \\
2040 & 439.950                                                              & 457.962                                                               & 1585.221               & 340.809               \\
2041 & 439.968                                                              & 459.408                                                               & 1567.070               & 341.122               \\
2042 & 439.910                                                              & 460.779                                                               & 1549.307               & 341.429               \\
2043 & 439.798                                                              & 462.074                                                               & 1531.982               & 341.731               \\
2044 & 439.631                                                              & 463.295                                                               & 1515.075               & 342.026               \\
2045 & 439.406                                                              & 464.441                                                               & 1498.566               & 342.315               \\
2046 & 439.123                                                              & 465.512                                                               & 1482.415               & 342.597               \\
2047 & 438.782                                                              & 466.510                                                               & 1466.602               & 342.873               \\
2048 & 438.383                                                              & 467.435                                                               & 1451.098               & 343.143               \\
2049 & 437.923                                                              & 468.286                                                               & 1435.903               & 343.407               \\
2050 & 437.404                                                              & 469.066                                                               & 1420.966               & 343.665                     \\ \hline \hline        
\end{tabular}
\end{table}

\begin{figure}[h]
\centering
\hspace*{-2.5cm}  
\includegraphics[width=1.25\textwidth]{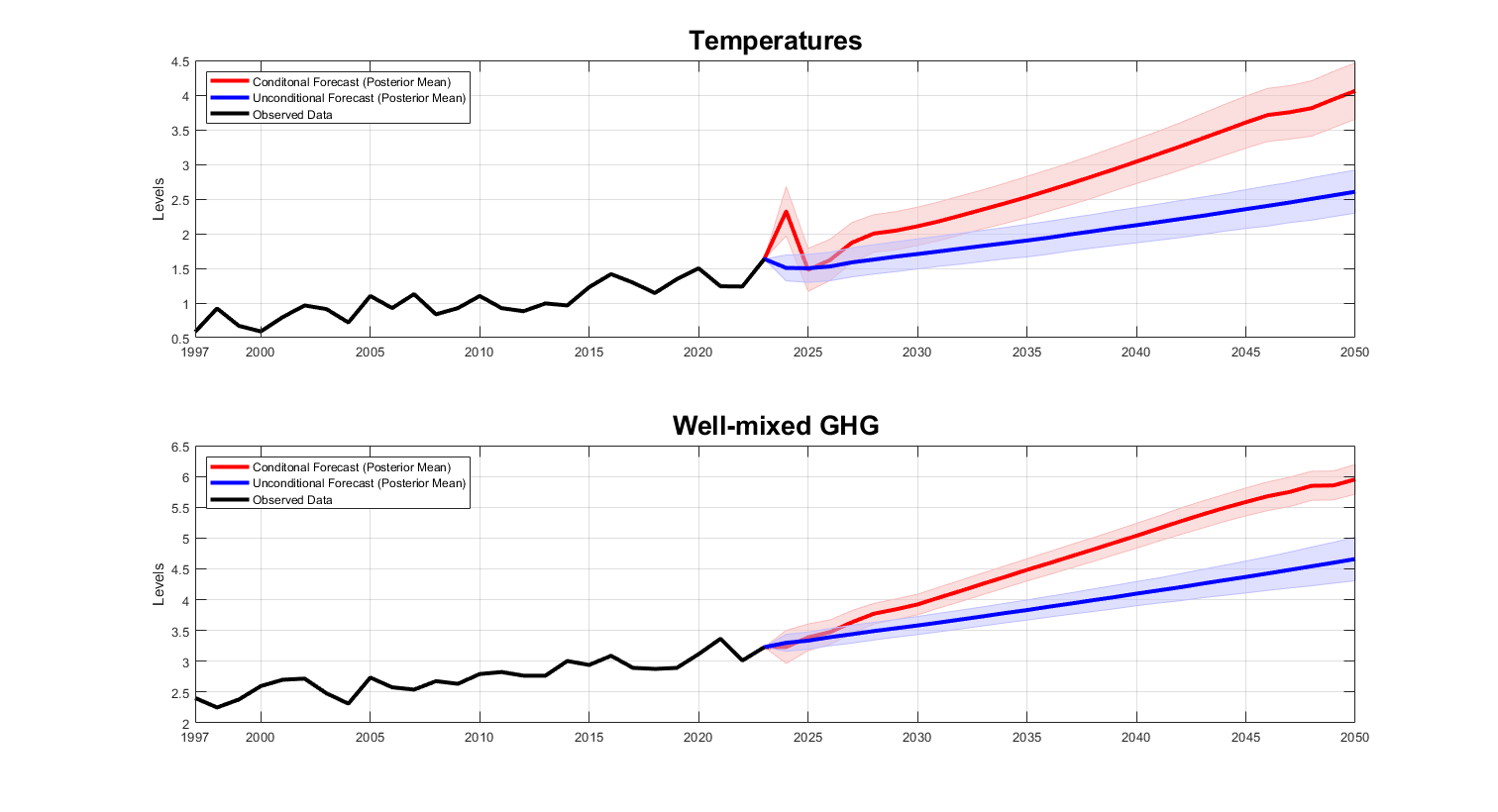}
\caption{\textit{Conditional and unconditional forecasts when $CO_2$, $CH_4$ and $N_2O$ emissions in 2024-2050 match the SSP adverse scenario projections. The shaded bands correspond to the $68\%$ coverage intervals while the solid black lines denote the in-sample values.}}
\label{Figure: Adverse Scenario MP}
\end{figure}

\begin{figure}[h]
\centering
\hspace*{-2.5cm}  
\includegraphics[width=1.25\textwidth]{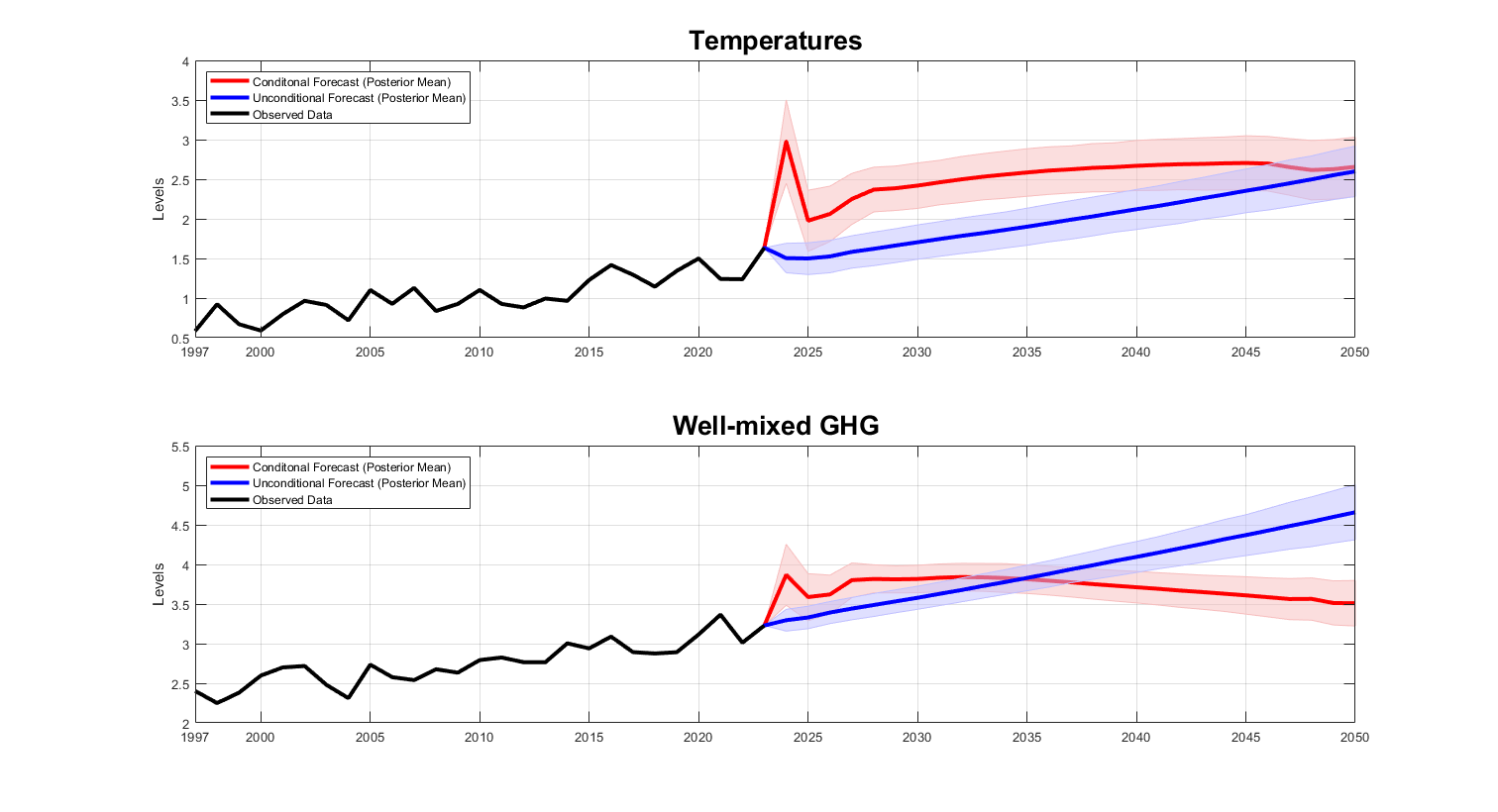}
\caption{\textit{Conditional and unconditional forecasts when $CO_2$, $CH_4$ and $N_2O$ emissions in 2024-2050 match the SSP optimistic scenario projections. The shaded bands correspond to the $68\%$ coverage intervals while the solid black lines denote the in-sample values.}}
\label{Figure: Optimistic Scenario MP}
\end{figure}

\bigskip

\end{document}